\documentclass[12pt]{iopart}
\usepackage{graphicx}
\usepackage{float}
\usepackage{cite}
\usepackage[usenames,dvipsnames]{color}
\begin{document}
\title[Tin-accompanied and true ternary fission of $^{242}$Pu] {Tin-accompanied and true ternary fission of
$^{242}$Pu}

\author{ M. Zadehrafi$^{1,2}$, M. R. Pahlavani$^{1}$, M. -R. Ioan$^{2}$}
\address {$^{1}$ Department of Physics, Faculty of Basic Science, University of Mazandaran, P.O.Box 47415-416, Babolsar, Iran.}
\address{$^{2}$ \textit{Horia Hulubei} National Institute for Physics and Nuclear Engineering (IFIN-HH),
P.O.Box MG-6, RO-077125, Bucharest-Magurele, Romania.}
\ead{zmastaneh@theory.nipne.ro}
\ead{m.pahlavani@umz.ac.ir}
\ead{razvan.ioan@nipne.ro}

\begin{abstract}
True ternary fission and Tin-accompanied ternary fission of $^{242}$Pu are studied by using `\textit{Three Cluster Model}'. True ternary fission is considered as formation of heavy fragments in the region $28\leq Z_1,Z_2,Z_3\leq 38$, with comparable masses. The possible fission channels are predicted from potential-energy calculations. Interaction potentials, Q-values and relative yields for all possible fragmentations in equatorial and collinear configurations are calculated and compared to each other. It is found out that ternary fission with formation of a double magic nucleus like $^{132}Sn$ is more probable than the other fragmentations. Also the kinetic energies of the fragments for the group $Z_1=32$, $Z_2=32$ and $Z_3=30$ are calculated for all combinations in the collinear geometry, as a sequential decay.

\end{abstract}

\vspace{2pc} \noindent{\it Keywords}: true ternary fission, three cluster
model, equatorial, collinear, kinetic energy.

\section{Introduction} \label{section.intro}
With the discovery of fission and its application to produce nuclear
energy at the late 1940s, most attempts of nuclear scientists had been focused logically on the
study of binary fission. Ternary fission was diagnosed as an
interesting source of high energy alpha particles. This rare type of
nuclear reaction was discovered by Chinese and French scientists \cite{Tsiang1946,Tsiang1947,Tsiangpr1947,San-Tsiang1947,T.San-Tsiang1947}.\\
Disintegration of
an unstable heavy/superheavy nucleus into three fission fragments, by ignoring
neutrons and other types of radiations is termed as cold ternary fission \cite{ref1,ref4,ref5,ref6,ref7,ref8,Rosen1950}. Ternary fission is an appropriate tool to study the behavior of a nuclear system at the scission point of a fissioning nucleus.\\
True ternary fission in which the
parent nucleus breaks up into three fragments of comparable (large) masses, occurs very rarely in some heavy/superheavy
nuclei with high fissility parameters \cite{plbNasirov2014,46}. This type of ternary fission has been studied
extensively in refs. \cite{Roy,Stoenner,Muga,MacMurdo,Kugler,Gottschalk,Grawert,Wu,Herbach,Pashkevich,Pyatkov2010}, but all its theoretical characteristics are not still well understood.\\
 By using double folding nuclear potential, a coplanar three-cluster approach was introduced to study the cold ternary fission of
$^{252}$Cf accompanied by $\alpha$ particle \cite{prc99Greiner}.
Poenaru \textit{et al}. \cite{Poenaru2000,Poenaru2003,Poenaru2005,1,2,3,5,6,7,8}
 have developed a macroscopic-microscopic model to study ternary
fission.\\
According to S\v{a}ndulescu \textit{et al}. \cite{29,31}, cold ternary fission can be considered like a process of cluster radioactivity; \textit{i.e.}, large number of nucleons are re-arranged in a cold process, from the ground state of the parent to the ground state of the three final products.\\
Without considering the cluster preformation factors, S\v{a}ndulescu \textit{et al}. \cite{32,33,34,35} have calculated the isotopic yields in the cold ternary fission of $^{248}$Cm, using double folding potential plus $M3Y$ nucleon-nucleon forces. Later, Florescu \textit{et al.} \cite{30} developed that model by including the preformation probability for $^{4}$He and $^{10}$Be accompanied ternary fission of the $^{252}$Cf isotope.\\
In the framework of the cluster picture and as an extension of the preformed cluster model (PCM) \cite{37}, `\textit{Three-Cluster Model}' (TCM) was introduced for studying the ternary fission process \cite{36}. This model has been used extensively to investigate the different theoretical aspects of ternary fission for various isotopes of Cf, U, Pu and Cm \cite{38,39,42,43,44,47}.\\
Despite the macroscopic approach of three-cluster model, its predictions and the obtained results are in good agreement with the available experimental data and other models \cite{Denisov,Greiner,Karpov,Oertzen,Adamian,Andreev,Nasirov2,R.B.Tashkhodjaev,R.B.Tashkhodjaev2}.\\
In our first publication \cite{Zadehrafi}, we studied the cold ternary fission of
the $^{250}$Cm by using three-cluster model in the equatorial geometry and
considering light charged particles as the fixed third fragment. In our second paper \cite{Zadehrafi2},
  a relatively heavy nucleus ($^{34}$Mg) is considered as the fixed third fragment to compare the results with light third fragments in ternary fission of $^{242}$Pu. The obtained results revealed that ternary fission of
 $^{242}$Pu accompanied by $^{34}$Mg, occurs with very low probability in the equatorial configuration.\\
In our recent investigation \cite{Zadehrafi3}, the equatorial and collinear configurations in the relatively heavy (A=14) accompanied ternary fission of $^{242}$Pu are compared. We also compared the results which are obtained using proximity and Yukawa plus exponential potentials as the nuclear part of the total potential, there.\\
In the present activity, we focus on the true ternary fission of the $^{242}$Pu isotope in a definite region of the mass and charge  for the three fragments. The new aspect of this study is the variation of both charge and mass numbers for all three fragments, it means expanding the area of the investigation and considering all possible combinations in a defined region for true ternary fission. Both equatorial and collinear geometries are considered, also kinetic energy of the fragments for the most favorable group of fragmentations is calculated.\\
  In Section \ref{section.method},
  a brief theoretical explanation of the TCM is presented.
The obtained results are presented and discussed in Section
\ref{section.result}. Finally, a summary of the present study
along with the concluding remarks is provided in Section
\ref{section.conclusion}.

\section{Theoretical framework} \label{section.method}
In the cold ternary fission, based on three-cluster model \cite{36}, the interaction potential of fragments is defined by
    \begin{equation}\label{eq.interaction potential}
      V=\sum_{i=1}^{3}\sum_{j>i}^{3}{(m^i_x+V_{Cij}+V_{Nij})}.
    \end{equation}
Here, $m^i_x$ are the mass excesses of three fragments in energy units, taken from standard mass tables \cite{mass-table3}. $V_{Cij}$ and $V_{Nij}$ are Coulomb and the nuclear potentials
 between each pair of the three interacting fragments, respectively.
The repulsive Coulomb potential between fragments $i$ and $j$, is as following
    \begin{equation}
      V_{Cij}=\frac{Z_i Z_j e^2}{C_{ij}},
    \end{equation}
where $Z_i$ and $Z_j$ are the charge numbers and $C_{ij}$ is the
distance between the centres of two fragments $i$ and $j$,
respectively.
    \begin{equation}
     {C_{ij}=C_i+C_j+s_{ij}}.
    \end{equation}
 Here, $C_i$ and $C_j$ are the S\"{u}ssmann central radii of the nuclei, and $s_{ij}$ is the distance between
 near surfaces of the nascent fragments $i$ and $j$. Note that $s=0$, $s>0$, and $s<0$ are correlated with `\textit{touching configuration}', `\textit{separated geometry}', and `\textit{overlap region}' of a pair of interacting nuclei, respectively.
 The S\"{u}ssmann radii are taken from ref. \cite{blocki1977}:
    \begin{equation}
    {C_x}={R_x}[1-(\frac{b}{R_x})^2],
    \end{equation}
where the subscript $x$ indicates the fragment number ($i$ and $j$= 1, 2 or 3), and
    \begin{equation}
      R_x=1.28A_x^{1/3}-0.76+0.8 A_x^{-1/3}
    \end{equation}
is the sharp radius of the fragment `$x$' with the mass number $A_x$. $b$ is the diffusivity parameter of
the nuclear surface (\textit{i.e.}, $b = \frac{\pi}{\sqrt{3}} a$ with $a=0.55 fm$) which
has been evaluated close to unity \cite{45}. Note that in TCM spherical shapes are considered for the decaying nucleus and all the fragments \cite{36}.\\
In the present research, the latest version of proximity nuclear potential (\textit{Prox2010}) \cite{45} has been used. According to this version of proximity potential, $V_{Nij}$ is defined as
    \begin{equation}
     V_{Nij}(s)=V_{Pij}(s)=4 \pi b \gamma  \overline{C} \Phi(\frac{s}{b}).
    \end{equation}
Here, $\gamma$ is the coefficient of nuclear surface tension, which is given by
    \begin{equation}
      \gamma=1.25284[1-2.345(N-Z)^2/A^2] \quad MeV/fm^2,
    \end{equation}
where $Z$, $N$, and $A$ are the proton, neutron, and mass numbers of
the compound system, respectively. The compound system means a nuclear system composed of a pair of fission products. \\
$\overline{C}$, the mean curvature radius, is evaluated as

    \begin{equation}
      \overline{C}=\frac{C_i C_j}{C_i+C_j}.
    \end{equation}

\begin{figure}[H]
\centerline{\includegraphics [width=5.5in, clip=true, trim=0 600 0 0] {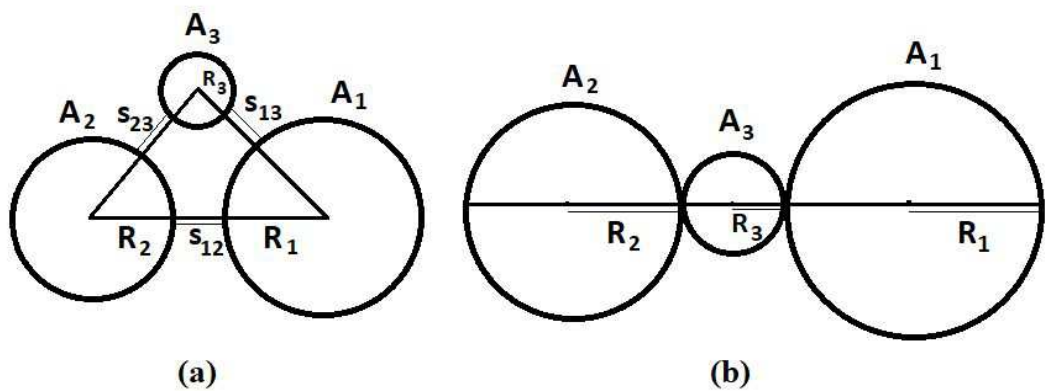}}
\caption{A scheme of the ternary fission fragments (a) equatorial configuration for separated fragments ($s>0$) (b) collinear touching configuration ($s=0$).}
\protect\label{fig.1}
\end{figure}

The universal function of proximity potential depends on the distance between each pair of fragments. This function is defined as follows
    \begin{equation}
        \Phi(\xi)=\Bigg\{
     \begin{array}{lc}
      -1.7817+0.9270 \xi+0.0169 \xi^2-0.05148 \xi^3 & for \quad 0\leq\xi\leq1.9475 \\
      -4.41 \exp(-\xi/0.7176) & for \quad \xi>1.9475.
      \end{array}
    \end{equation}
Here, $\xi=s/b$ is a function of the distance between interacting nuclei.
It is assumed that in the equatorial configuration, the three fission products are separated symmetrically
  from each other with the same speed.
  Therefore, one can consider the equal separation
  distances between each pair of tripartition fragments; \textit{i.e.}, $s=s_{12}=s_{13}=s_{23}$.
  In fact, the lightest fragment moves faster than two heavier ones, due to the repulsive Coulomb forces. If $A_3$ has been considered as the lightest fragment, the relation between separation distances is $k\times s_{12}=s_{13}=s_{23}$, with $0 < k \leq 1$. However, it is shown in the ref. \cite{36} that the trends of relative yields and fragmentation potential barriers  are not affected by the $k$-value and one can consider $k=1$, as a reliable assumption.
  \\On the other hand, in the collinear configuration with $A_3$ in the middle, the surface distance
  between fragments 1 and 3 or 2 and 3 is $s=s_{13}=s_{23}$. But for the fragments 1 and 2, this parameter is written as
  \begin{equation}
      s_{12}=2(C_3+s),
  \end{equation}
 which in both geometries, $s=0$ corresponds to the touching configuration. The Q-value of the cold ternary fission is given by
    \begin{equation}\label{eq.interaction energy}
      Q=M-\sum_{i=1}^{3}{m_i},
    \end{equation}
which it should be positive to make the spontaneous reaction possible. $M$ is the mass excess of the decaying nucleus and $m_i$ are
the mass excesses of the fission products in the energy unit. Also, since the parent and all the fragments are considered in their ground state, Q-value appears as the kinetic energy of the three fragments and can be defined as $Q = E_1 + E_2 + E_3$ with $E_i (i=1,2,3)$.\\
The relative yield for each fragmentation channel is calculated using
    \begin{equation}
        Y(A_i,Z_i)=\frac{P(A_i,Z_i)}{\sum{P(A_i,Z_i)}},
    \end{equation}
where $P(A_i, Z_i)$ is the penetrability of the $i$-th fragment through the three-body potential barrier. The one-dimensional $W.K.B$ approximation is used to calculate the probability of penetration through the potential barrier \cite{36},
    \begin{equation}
      P=\exp\left\{-\frac{2}{\hbar}\int^{s_2}_{s_1}{\sqrt{2\mu(V-Q)}ds}\right\}.
    \end{equation}
The touching configuration has been chosen as the first turning point $s_1=0$,
and the second turning point $s_2$ should satisfy the $V(s_2)=Q$ equation, in the above integral.\\
The reduced mass of the three fission products is defined as
    \begin{equation}
    \mu=m (\frac{A_1 A_2 A_3}{A_1 A_2+A_1 A_3+A_2 A_3}),
    \end{equation}
where $m$ is the average mass of the nucleon and $A_1$, $A_2$, and
$A_3$ are the mass numbers of the three fragments.\\
A scheme of the tripartition fragments in equatorial and collinear geometries is shown in fig. \ref{fig.1}. Touching configuration in this figure ($s=0$), is related to the first turning point in the integral of the equation (13).

\section{Results and discussions} \label{section.result}
In the first step of the study of true ternary fission in the $^{242}$Pu isotope, all possible fragmentations with $28\leq Z \leq 38$ are extracted. The imposed condition in this research is $Z_3\leq Z_2 \leq Z_1$, for avoiding of the repetition of the fragment
arrangements in the calculation of the potential energies. Considering this condition, 14 groups of fragments with various atomic numbers are selected.\\ In the second step, for each group, all possible combinations with different mass numbers are listed. So in each one of those 14 groups, about 300 subgroups have been identified. Then the interaction potentials, Q-values, penetration probabilities and relative yields have been calculated for each individual fragmentation in the collinear (with the lightest fragment in the middle of the arrangement) and equatorial geometries. Note that interaction potentials are calculated in touching-fragments configuration.\\ Due to the huge amount of data, presentation of all calculated results is virtually impossible. Therefore, to be able to compare the results, the minimum of potentials has been chosen from each category.\\ Q-values and minimum interaction potentials in the collinear and equatorial geometries are presented in the table 1. As it is simply evident from this table, in this region of mass and charge numbers, the potential barriers of collinear configurations are lower than the equatorial ones. This result has been verified with the results presented in the refs. \cite{39,42,46}. Also in the most of the combination groups, there is at least one fragment with neutron and/or proton closed shell (bold numbers in the table 1).\\

\begin{figure}[hp]
  {\footnotesize{\textbf{Table 1.} Q-values and minimum interaction potentials for 14 groups of $Z_1$, $Z_2$ and $Z_3$ between 28 and 38, with the condition $Z_3\leq Z_2\leq Z_1$. 7 highlighted groups are represented in the Fig. \ref{Z_V} for more visual comparison}}
 \centerline{\includegraphics[clip=true, trim=0 50 0 50]{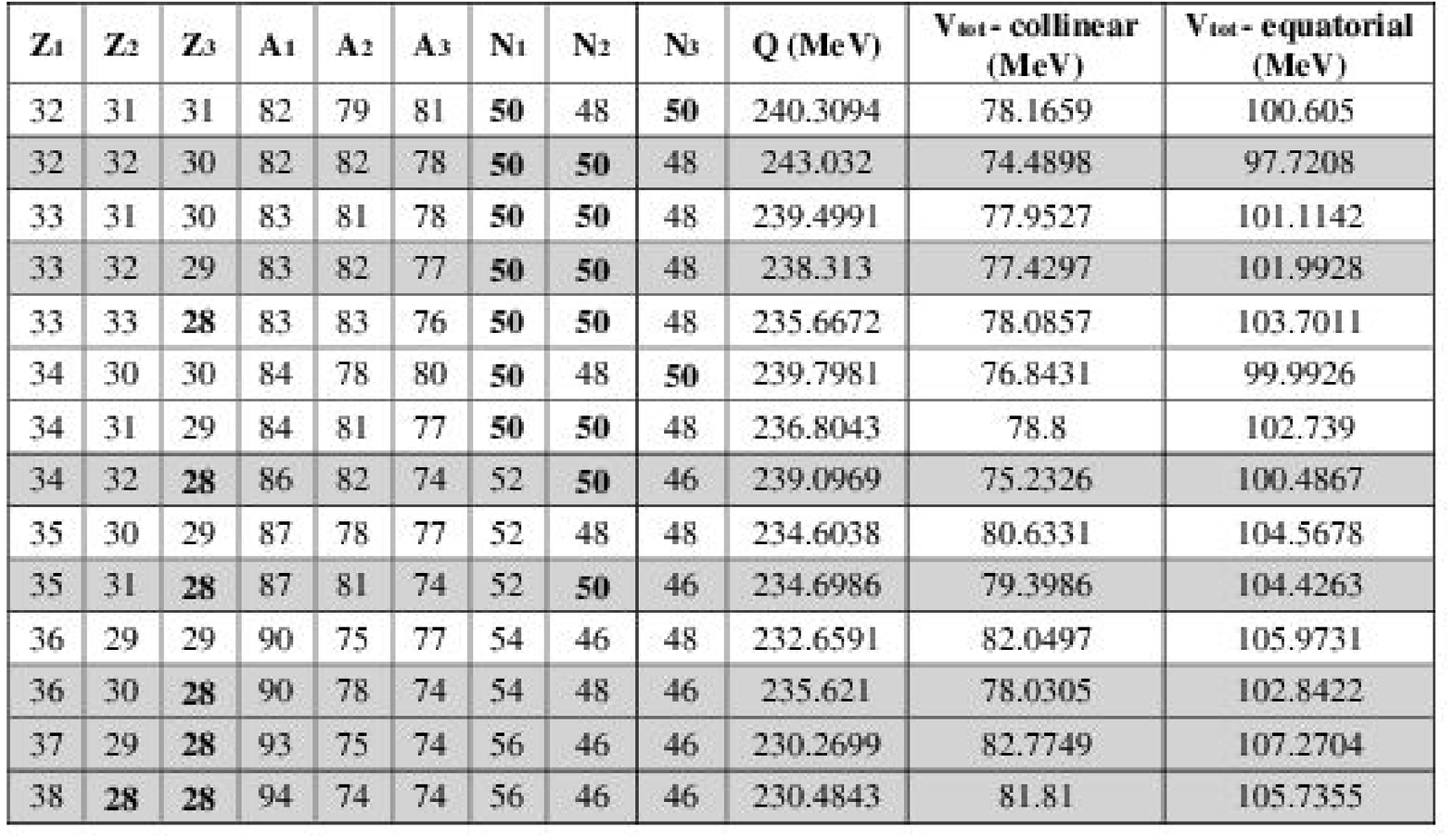}}
\end{figure}

\begin{figure}[H]
\centerline{\includegraphics[width=4.5in]{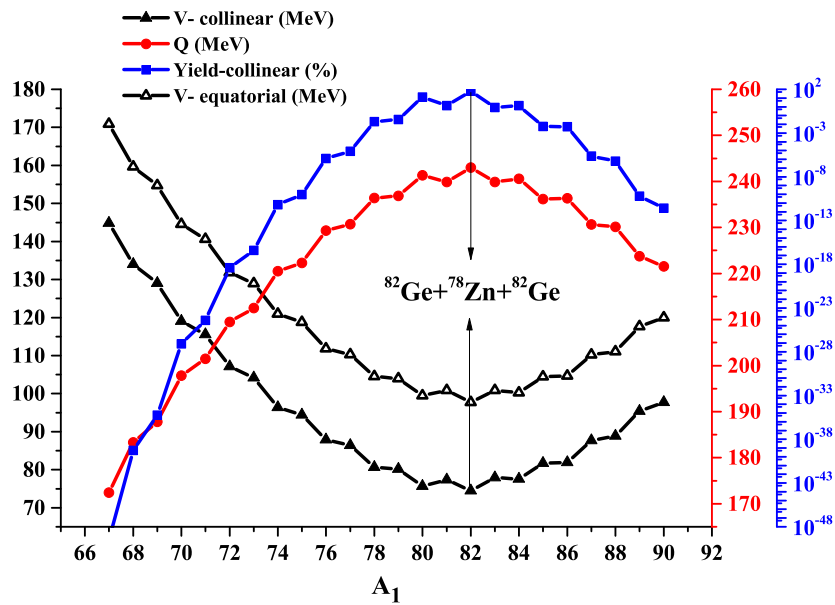}}
      \caption{Interaction potentials in the collinear and equatorial configurations (associated with the left vertical axis), Q-values (associated with the right vertical axis) and relative yields in the collinear geometry (associated with the logarithmic axis) for the combinations with $Z_1=32$, $Z_2=32$,  $Z_3=30$ and different mass numbers, plotted as a function of $A_1$.
\protect\label{sheet3}}
\end{figure}

 In the group $Z_1=32$, $Z_2=32$, and $Z_3=30$ which has the lowest minimum interaction potential among all those 14 groups, the most favorable combinations with the same $A_1$ have been chosen and the variations of the interacting potentials, Q-values and relative yields are plotted as a function of fragment mass number $A_1$. The results are presented in fig. \ref{sheet3}. Note that three vertical axes in this figure have been scaled differently.\\

\begin{figure}[H]
\centerline{\includegraphics[width=4.5in]{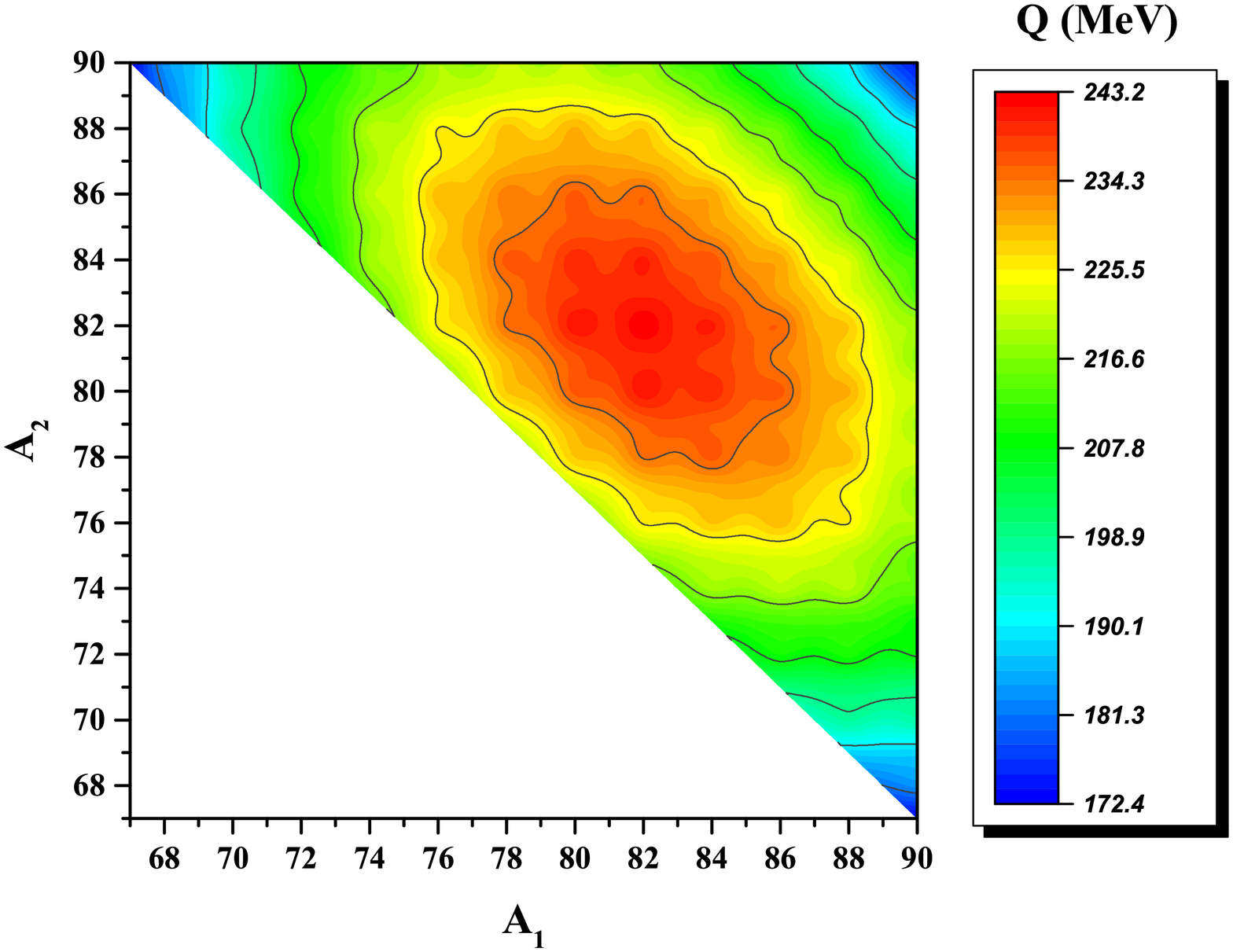}}
      \caption{Contour map of the Q-values for all possible combinations, for the breakup $^{242}Pu\rightarrow ^{A_1}Ge+ ^{A_3}Zn+ ^{A_2}Ge$, plotted as a function of fragment mass numbers $A_1$ and $A_2$.
\protect\label{sheet3_Q}}
\end{figure}

\begin{figure}[H]
\centerline{\includegraphics[width=4.5in]{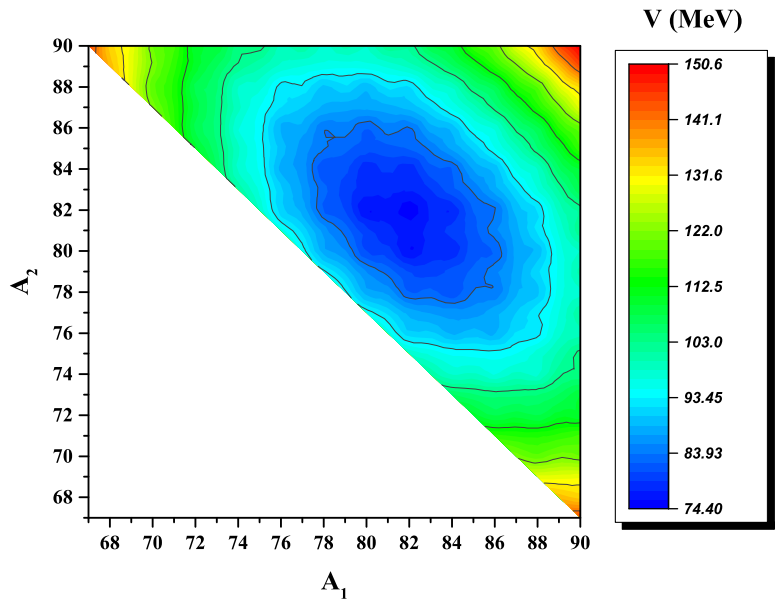}}
      \caption{Contour map of the interaction potentials (collinear geometry) for all possible combinations, for the breakup $^{242}Pu\rightarrow ^{A_1}Ge+ ^{A_3}Zn+ ^{A_2}Ge$, plotted as a function of fragment mass numbers $A_1$ and $A_2$.
\protect\label{sheet3_V}}
\end{figure}

From the fig. \ref{sheet3}, it is obvious that increasing of the Q-value and relative yield is equivalent to decrease of the interaction potential and vise versa. However, this equivalency is not always valid. According to the fig. \ref{sheet3}, $Z$-values are constant in all considered combinations. If both $A$ and $Z$ values vary among different combinations, one may see that there is no specific relation between Q-values and relative yields or interaction potentials (see subsection \ref{subsection}).\\
In the fig. \ref{sheet3}, the minimum of the interaction potentials and maximum of the yields and Q-values belongs to the combination $^{82}Ge+^{78}Zn+^{82}Ge$ with magic neutron numbers for two Ge isotopes ($N=50$). Also for this group ($Z_1=32$, $Z_2=32$, and $Z_3=30$), the contour maps are represented considering all 300 possible combinations with various mass numbers. It can be seen that maxima of the Q-values (fig. \ref{sheet3_Q}) which correspond to minima of the interaction potentials (fig. \ref{sheet3_V}) belong to a region that mass numbers of $A_1$ and $A_2$ (and consequently $A_3$) are close together, and this region can be considered as the region of true ternary fission.\\

By more scrutiny of table 1, one can conclude that: (1) Even-mass fragments possess lower potential barriers than the odd-mass ones (in agreement with \cite{38,39,santhosh,santhosh2}). (2) Neutron closed shell structures are more important than the proton closed shells, for lowering the potential barrier (compatible with \cite{36,Ismail}). (3) The closed shell structure of the heaviest fragment plays a key role for the produced more favorable channels (in agreement with \cite{Ismail}). (4) Fragments with less difference in mass numbers, contain lower potential barriers and higher Q-values, comparing to other fragmentations (upper and lower rows of the table 1).

\subsection{Comparison between true and Tin-accompanied ternary fission of  $^{242}Pu$} \label{subsection}
\begin{figure}[H]
\centerline{\includegraphics[width=4.5in]{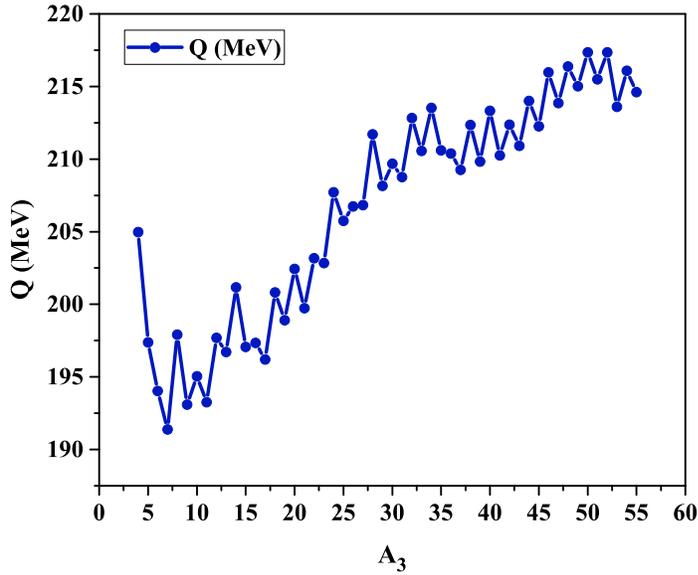}}
\caption{Q-values of the breakup $^{242}Pu\rightarrow ^{132}Sn+ A_3+ A_2$.}
\protect\label{Q_value}
\end{figure}

\begin{figure}[H]
\centerline{\includegraphics[width=4.5in]{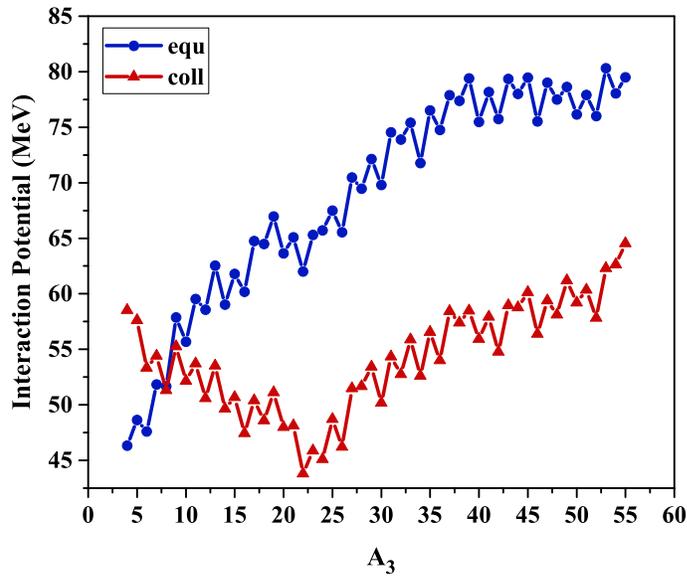}}
      \caption{Charge minimized interaction potentials, for the breakup\\ $^{242}Pu\rightarrow ^{132}Sn+ A_3+ A_2$, in the collinear and equatorial geometries.
\protect\label{Sn-equ-coll}}
\end{figure}

In this part of study, we consider a double magic nucleus ($^{132}Sn$) as the fixed fragment, in order to compare with the previous results of true ternary fission of $^{242}Pu$. Like the previous section, all possible ternary channels are considered. Then Q-values and charge minimized potentials in the equatorial and collinear configurations have been calculated and plotted as a function of $A_3$ (lightest fragment) in figs. \ref{Q_value} and \ref{Sn-equ-coll}, respectively. As it is clear from these figures, there is no specific relation between Q-values and interaction potentials, due to variation of both $A$ and $Z$ values. In fact, the actual possibility of ternary fission is decided by the potential barrier properties and not by the
released energy.\\
It can be seen in the fig. \ref{Sn-equ-coll} that collinear geometry possess the lower potential barrier than the equatorial geometry, except for very light third fragments. The lowest barrier in collinear geometry belongs to the combination $^{132}Sn+^{22}O+^{88}Kr$. Very similar results for ternary fission of $^{252}Cf$ have been reported in ref. \cite{Denisov}.\\

\begin{figure}[H]
\centerline{\includegraphics[width=4.5in]{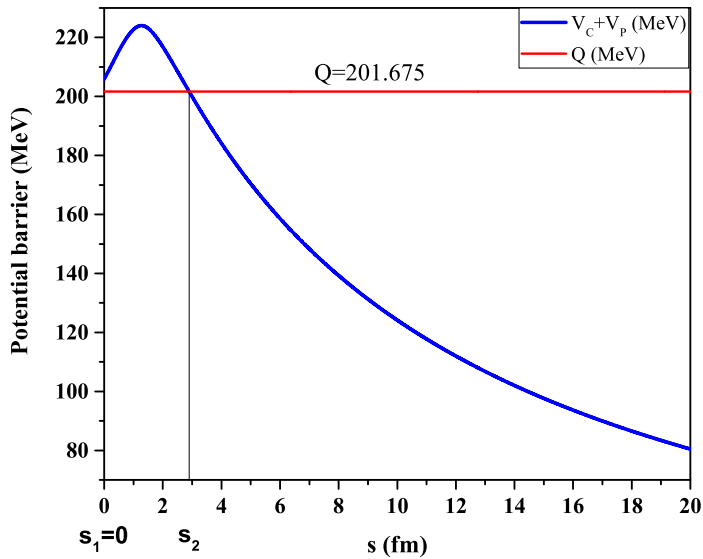}}
      \caption{The potential barrier $(V_C+V_P)$ as a function of separation parameter $s$, for the breakup $^{242}Pu\rightarrow ^{132}Sn+ ^{22}O+ ^{88}Kr$. The turning points and Q value are also labeled.
\protect\label{V_s}}
\end{figure}

Variation of the potential barrier ($V_C+V_P$) as a function of separation parameter ($s$) is presented in fig. \ref{V_s}, for the combination $^{132}Sn+ ^{22}O+ ^{88}Kr$. The potential is calculated by varying the value of $s$ uniformly, from the touching point till beyond. It is to be mentioned here that the potentials of overlap region are not favored in this model. Indeed, shifting the first turning point from touching configuration $(s_1=0)$, to a point like $s_0$ $(V(s_0)=Q)$, will lead to the model of Shi and Swiatecki (Ref. \cite{shi}) for penetrability calculations. More information about calculation of penetrability by using the two turning points are represented in Refs. \cite{37}, \cite{36}, \cite{Zadehrafi} and \cite{Zadehrafi2}.

\begin{figure}[H]
\centerline{\includegraphics[width=4.5in]{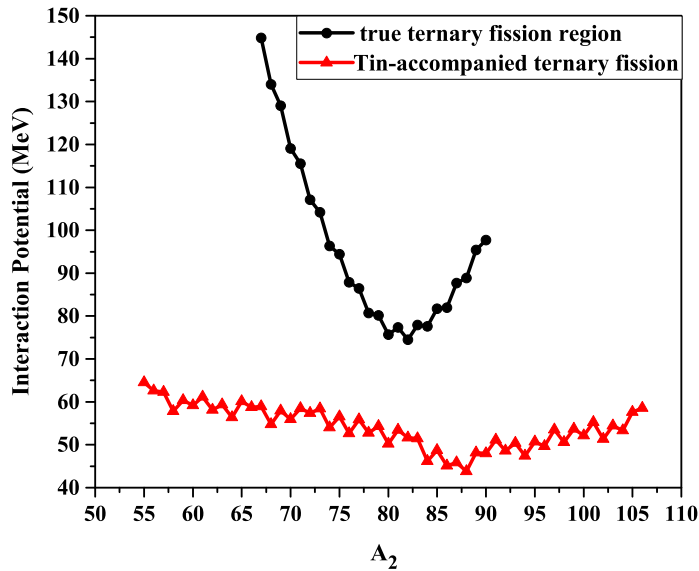}}
      \caption{Comparison of potential barriers for true ternary fission region and Tin-accompanied ternary fission of $^{242}Pu$ in the collinear configuration.
\protect\label{Sn_Vs_32}}
\end{figure}

In fig. \ref{Sn_Vs_32}, the interaction potentials for true ternary fission region ($Z_1=32$, $Z_2=32$, $Z_3=30$) and Tin-accompanied ternary fission of  $^{242}Pu$ are compared. It is obvious from this figure that in collinear configuration, ternary potential barriers with $^{132}Sn$ as the fixed fragment are much lower than the other group. Since $^{132}Sn$ is a double magic isotope ($Z=50$ and $N=82$), this result emphasizes the importance of the closed shell structures in production of favorable ternary channels.\\

\begin{figure}[H]
\centerline{\includegraphics[width=4.5in]{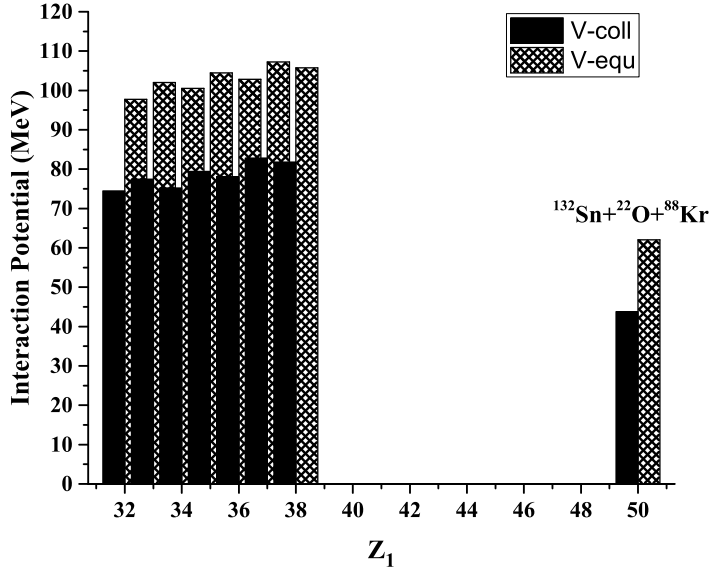}}
      \caption{Comparison of minimum interaction potentials for true and Tin-accompanied ternary fission of $^{242}Pu$ in the collinear and equatorial geometries.
\protect\label{Z_V}}
\end{figure}

In order to have a more visual comparison, 7 groups with different $Z_1$ (highlighted in the table 1) are picked and shown in fig. \ref{Z_V} as bar graphs. The combination $^{132}Sn+^{22}O+^{88}Kr$ is also shown in this figure. It is evident that in those 7 groups, there is no significant difference between the magnitudes of interaction potentials for fragments with various $Z$ (less than 10 $MeV$). But ternary fragmentation potential barrier with $^{132}Sn$ as the fixed fragment is almost $30$ $MeV$ lower than the others.\\

\subsection{Kinetic energy of the fragments in the group $Z_1=32$, $Z_2=32$, and $Z_3=30$}
\begin{figure}[H]
\centerline{\includegraphics[width=4.5in]{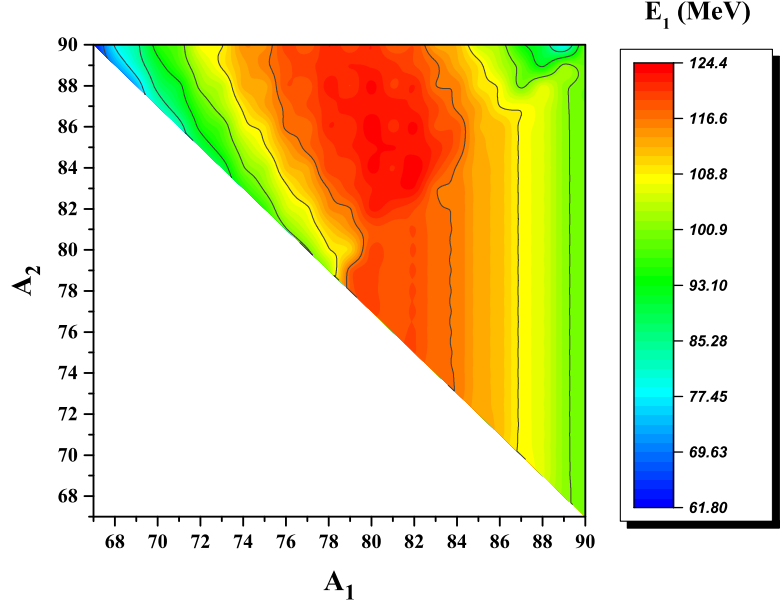}}
      \caption{Kinetic energies of the fragment $^{A_1}Ge$ as a function of $A_1$ and $A_2$, for the collinear breakup $^{242}Pu\rightarrow ^{A_1}Ge+ ^{A_3}Zn+ ^{A_2}Ge$.
\protect\label{E1}}
\end{figure}

\begin{figure}[H]
\centerline{\includegraphics[width=4.5in]{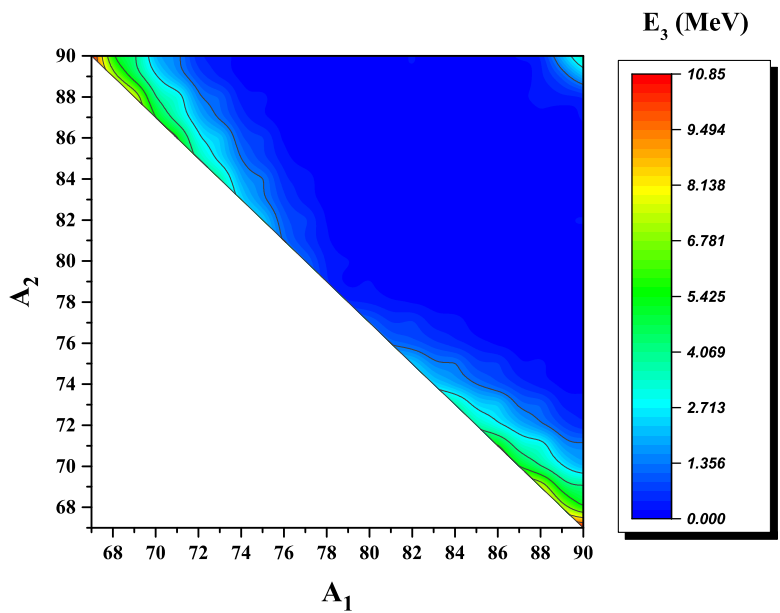}}
      \caption{Kinetic energies of the fragment $^{A_3}Zn$ as a function of $A_1$ and $A_2$, for the collinear breakup $^{242}Pu\rightarrow ^{A_1}Ge+ ^{A_3}Zn+ ^{A_2}Ge$.
\protect\label{E3}}
\end{figure}

In order to calculate the kinetic energies of the fragments, we concentrate on the decay $^{242}Pu\rightarrow ^{A_1}Ge+ ^{A_3}Zn+ ^{A_2}Ge$, which possess the lowest potential barrier among those 14 groups. Also the collinear tripartition is considered as a sequential decay, which means that the ternary fragmentation happens in two steps.  In the first step, the unstable parent nucleus with mass number $A$ breaks into fragments $A_i$ and $A_{jk}$. Then in the next step, the
composite fragment $A_{jk}$ fissions into fragments $A_j$ and $A_k$. In this study $i$, $j$ and $k$ are referred to fragment numbers 1, 3, and 2, respectively. We assume that, in both steps, energy and momentum
of the system are conserved. In order to calculate the kinetic energy, we employ the method presented in ref. \cite{47}. The mathematical method of calculation for the kinetic energy is presented here briefly. For more details, the interested readers can  return to the ref. \cite{47}
    \begin{equation} \label{1}
        Q_I=M_x(A)-[m_x(A_1)+m_x(A_{23})].
    \end{equation}
    \begin{equation} \label{2}
        Q_{II}=m_x(A_{23})-[m_x(A_2)+m_x(A_3)].
    \end{equation}
Equations (\ref{1}) and (\ref{2}) are related to the steps one and two, respectively. $M_x$ is the mass excess of the parent and $m_x$ are the mass excesses of the fragments in each step.\\ At the first step, velocity of the composite nucleus is obtained using
    \begin{equation}
        v_{23}=+\sqrt{(\frac{2m_1}{m_1+m_{23}})(\frac{Q_I}{m_{23}})},
    \end{equation}
and similarly, the velocity of the fragment 1 is as follows
    \begin{equation}
        v_1=-\sqrt{(\frac{2m_{23}}{m_1+m_{23}})(\frac{Q_I}{m_1})}.
    \end{equation}
Here, $m$ are the masses of the mentioned fragments, expressed in the energy unit.\\ For the velocities of fragments 2 and 3 in the second step, we have
    \begin{equation}\label{3}
      v_2=\frac{m_2 m_{23} v_{23}\pm\sqrt{\zeta^2}}{m_2^2+m_2 m_3},
    \end{equation}
where
    \begin{equation}\label{4}
      \zeta^2=m_2^2 m_{23}^2 v_{23}^2-[(m_2^2+m_2 m_3)\times(m_{23}^2 v_{23}^2- 2 m_3 Q_{II}- m_3 m_{23} v_{23}^2)].
    \end{equation}
    \begin{equation}\label{5}
      v_3=-[\frac{m_2 v_2- m_{23} v_{23}}{m_3}].
    \end{equation}
Finally, using well known formula $E=\frac{1}{2} m v^2$, the kinetic energies of all three fragments are obtained.\\
The kinetic energies of the $^{A_1}Ge$ and $^{A_3}Zn$ fragments are presented in figs. \ref{E1} and \ref{E3} as a function of $A_1$ and $A_2$ for all 300 combinations.\\

\begin{figure}[H]
\centerline{\includegraphics[width=4.5in]{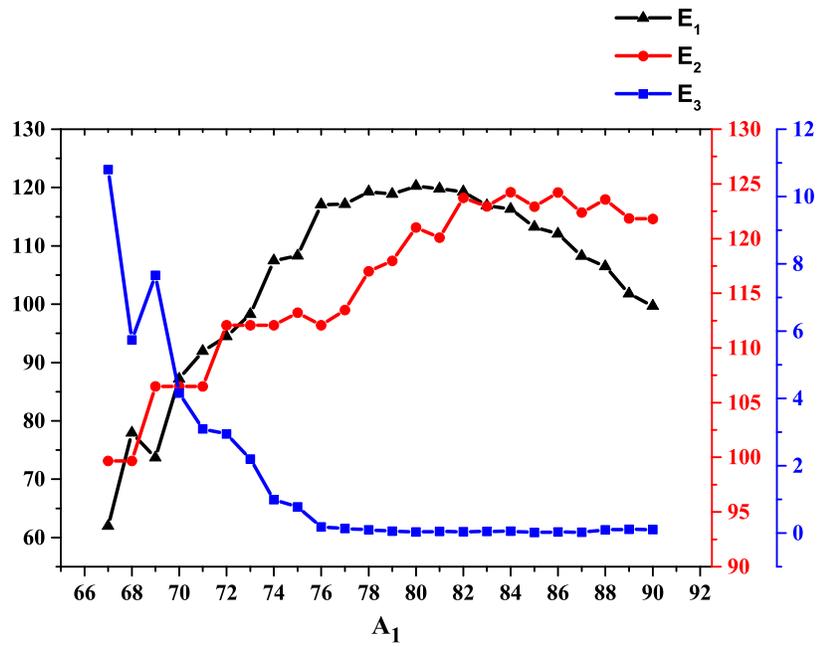}}
      \caption{Kinetic energies of the fragments $A_1$, $A_2$, and $A_3$ for sequential collinear decay $^{242}Pu\rightarrow ^{A_1}Ge+ ^{A_3}Zn+ ^{A_2}Ge$. Vertical axes from left to right, are correlated with $E_1$, $E_2$, and $E_3$, respectively. It is clear that the fragment number 3 is almost at rest.
\protect\label{2D-kinetic}}
\end{figure}

 As it is clear from fig. \ref{E3}, the light fragment that is located in the middle of the collinear arrangement, takes a very small part of the total kinetic energy and the rest of the total kinetic energy is removed by other two fragments. This fact about third fragment, can be the reason of its escaping from experimental detection. This result is in agreement with the ref. \cite{47}.\\
The kinetic energies of the fragments for the combinations which are mentioned in fig. \ref{sheet3} are presented as a two dimensional graph in fig. \ref{2D-kinetic}. The relative yields, Q-values and total kinetic energies of this group are also listed in table 2. One may see that Q-values and total kinetic energies for each fragmentation in the table 2  are almost equal. This result is due to considering ternary fissions as a cold process.\\

\begin{figure}[H]
  {\footnotesize{\textbf{Table 2.} Some calculated data for minimized potential combinations of the breakup $^{242}Pu\rightarrow ^{A_1}Ge+ ^{A_3}Zn+ ^{A_2}Ge$. for each value of $A_1$, the interaction potential is minimized. Therefore, 24 combinations among 300 are chosen (Yields less than $10^{-7}$ are denoted as ``0'')}}
 \centerline{\includegraphics[clip=true, trim=0 50 0 50]{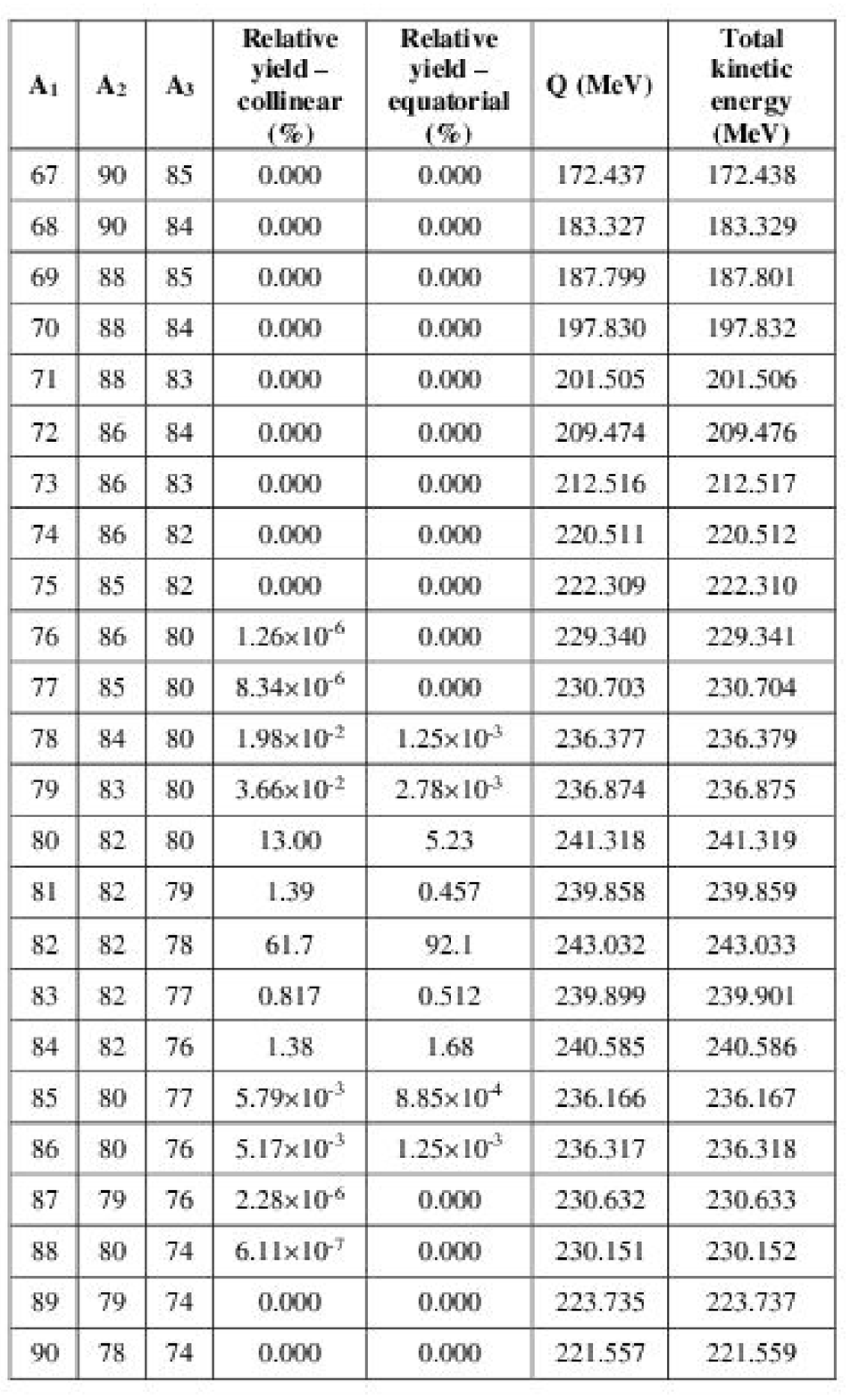}}
\end{figure}

\section{Conclusion} \label{section.conclusion}
True ternary fission of the $^{242}Pu$ isotope is studied and compared with the Tin-accompanied ternary fission of the parent. The most probable ternary fission path is predicted as the one which has a minimum in the interaction potential, with respect to the mass and charge asymmetries. The obtained results reveal that in the presented region of mass and charge numbers ($28\leq Z_1,Z_2,Z_3\leq 38$), the collinear geometry is preferred to the equatorial geometry. Also closed shell structures play an inevitable role in the potential barrier height. Indeed, closed neutron shells are more effective in lowering the potential barrier than the closed proton shells.
The most favorable combinations belong to a region where the fragments have comparable mass and charge numbers.\\ However, comparing with Tin-accompanied ternary fission of $^{242}Pu$, it is found out that this channel is more favorable than the true ternary fission region. This fact can be interpreted in two ways: (1) $^{132}Sn$ is a magic nucleus, both in proton and neutron numbers. (2) Three fragments with comparable sizes (true ternary fission) are less likely to be appeared in the exit channel.\\
The kinetic energies of the fragments in the group $Z_1=32$, $Z_2=32$, and $Z_3=30$ are also calculated in the collinear geometry as a sequential decay. The obtained results show that the most part of the total kinetic energy is carried by two $Ge$ nuclei, and $Zn$ as the middle part of the arrangement won't have considerable kinetic energy. This fact can be the reason of escaping from experimental detection, by the third fragment.


\end{document}